# SpareCodeSearch: Searching for Code Context When You Have No Spare GPU

Minh Nguyen
*School of Computer Science*
*University College Dublin*
Dublin, Ireland
minh.a.nguyen@ucdconnect.ie

*Abstract*—Retrieval-Augmented Generation (RAG) frameworks aim to enhance Code Language Models (CLMs) by including another module for retrieving relevant context to construct the input prompt. However, these retrieval modules commonly use semantic search, requiring substantial computational resources for training and hosting these embedded models, making them infeasible to integrate into lightweight applications such as in-IDE AI-based code completion. In this solution paper, we prove that using keyword-search is sufficient to retrieve relevant and useful code context inside large codebases, without the need for extensive GPU resources. The usefulness of code contexts found by our solution is demonstrated through their completion results on the Code Context Competition's benchmark, reaching 0.748 and 0.725 chRF scores on Kotlin and Python tracks, respectively.

*Index Terms*—Code Context, Code Search, Code Language Models, Code Completion, Keyword-based Search

## I. Motivation

Code Language Models (CLMs) have shown great promise in generating code, given the right contexts in their input prompts [1], [2]. Our solution - **SpareCodeSearch** - is developed with a philosophy that emphasizes simplicity and efficiency, focusing on the question of how to develop a code search tool that is most suitable for the task of automated code completion. This means that the tool must satisfy the requirements of lightweight deployment and fast retrieval times, easily integrate into existing CLMs, and, at the same time, be easily extensible to multiple programming languages. Semantic code search, while it has been proven effective in retrieving code snippets from natural language queries [3], is not suitable for these requirements, due to its reliance on resource-intensive language models to generate vector embeddings [4], which cannot be easily deployed in resource-constrained environments such as IDEs. In those scenarios, indexing a large codebase could take a significant amount of time and resources, making it impractical to combine with the computationally demanding LLM-based code completion. The learning-based nature of this method requires retraining and fine-tuning to mitigate knowledge cut-off, causing it to perform differently across different programming languages [3], [5].

We believe that a keyword-based search approach can effectively address these challenges. It has been shown that most developers when searching code, look for specific code examples or patterns that match their pre-existing knowledge and familiarity [6]. This tendency gives rise to the development of IDE features such as "Go to Definition" or "Search Everywhere"[1] which are heavily reliant on user-provided keywords. Coming back to the Code Context Competition, we extend the original question of "How to find relevant code context from a codebase that helps language models generate better code completion" into "How to effectively integrate existing keyword-based code search engines into automated code completion systems"

## II. Technical Details

### A. System Architecture

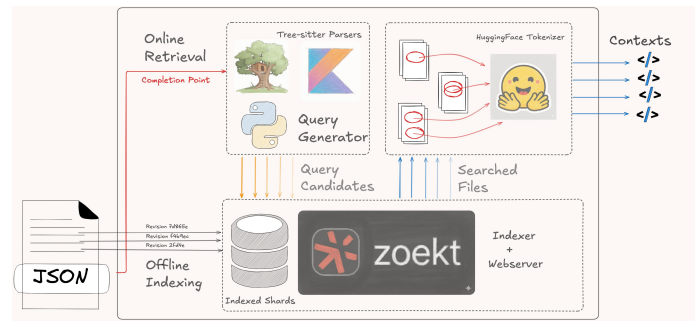

Fig. 1: System architecture overview of SpareCodeSearch.

The code search engine used in our solution is Zoekt[2]. Zoekt is an open-source fast, scalable code search engine that is designed to handle large codebases efficiently. Zoekt has been adapted by Sourcegraph and Gitlab as their backend for indexed code search [7]. Sourcegraph even integrated it into their own conversation-based AI Coding Assistant Cody [8]. Zoekt also shares the similar query language with other propietary code search engine such as Github Code Search [9] and Google Code Search [10], making our solution extensible to these engines in the future. Zoekt also has a separate index server and web server, allowing it to be deployed in microservices architectures, making it easy to integrate and customize with other systems.

---

[1]https://www.jetbrains.com/help/idea/searching-everywhere.html
[2]https://github.com/sourcegraph/zoekt

Looking at Figure 1, our system can be divided into two main phases: Offline Indexing and Online Retrieval

*B. Offline Indexing Phase with Zoekt*

With the provided public Kotlin and Python sets provided by the competition, we build an index using Zoekt. For each datapoint in the original JSONL file, Zoekt indexer server looks at its corresponding revisions, and creates an index shard. Each shard is a compressed representation of the codebase at that specific revision, optimized for fast keyword-based search [10]. The indexing process requires *ctags*[3] for extracting and saving symbols from the codebase to the index. For Kotlin public dataset, there are a total of 400 shards, belonging to 19 unique repositories. For Python, the total number of shards is 247, indexed from 20 repositories. See Figure 2 for the distribution of the number of revisions per unique repository. We build our index locally, on a Macbook M3 Air with 16GB ram, using Docker containers to ensure a consistent and isolated environment. The indexing process took approximately 10 minutes for each dataset, with the created shards being mounted to a Docker volume. After indexing, Zoekt web server will be turned on to serve the indexed shards, and expose its **/search** JSON API for retrieval.

| Language | Number of Revisions (Completion point) | Number of unique repos |
|---|---|---|
| Kotlin | 400 | 19 |
| Python | 247 | 20 |

TABLE I: Number of revisions per unique repositories for Kotlin and Python in public sets.

*C. Online Context Retrieving Phase*

The Online Retrieval phase begins after the initial indexing is complete, and the Zoekt web server is running. This phase involves two main modules: the Zoekt Query Generator and the Post-processor of search results.

| Query Variation | Kotlin | Python |
|---|---|---|
| functions_classes_naive | 371 | 237 |
| functions_classes_or | 371 | 237 |
| functions_classes_top5 | 214 | 201 |
| functions_classes_top4 | 250 | 209 |
| functions_classes_top3 | 275 | 218 |
| functions_classes_regex | 337 | 230 |
| navigation_naive | 346 | 219 |
| navigation_unpacked | 346 | 219 |
| navigation_unpacked_or | 346 | 219 |
| navigation_unpacked_top5 | 293 | 154 |
| navigation_unpacked_top4 | 307 | 176 |
| navigation_unpacked_top3 | 320 | 191 |
| navigation_regex | 346 | 219 |
| identifiers_naive | 396 | 244 |
| identifiers_or | 396 | 244 |
| identifiers_top5 | 386 | 226 |
| identifiers_top4 | 387 | 230 |
| identifiers_top3 | 388 | 236 |
| identifiers_regex | 396 | 244 |

TABLE II: Number of queries generated for each query construction strategy, for Kotlin and Python datasets respectively.

[3]https://github.com/universal-ctags/ctags

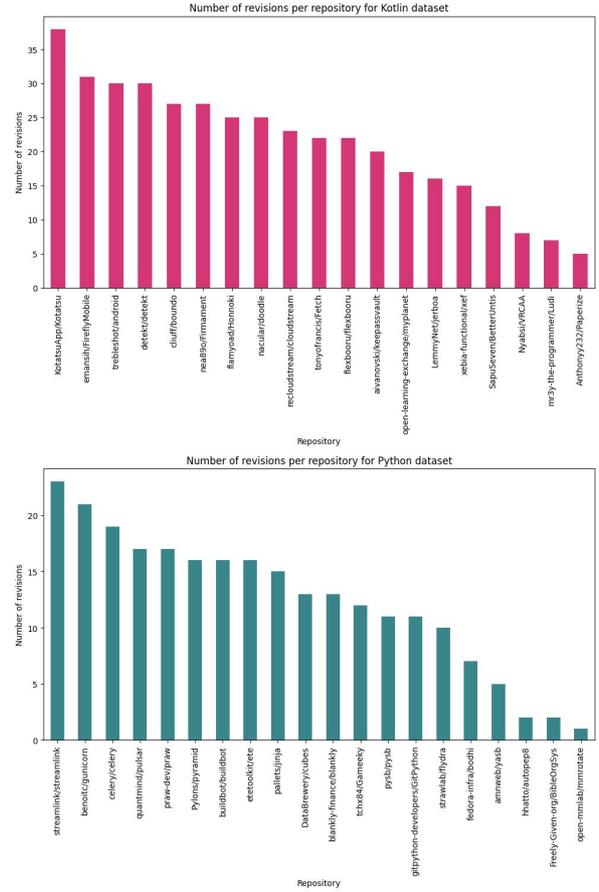

Fig. 2: Number of revisions for each unique repository indexed by Zoekt. (Public set only)

*1) Zoekt Query Construction Strategies:* For every original completion point, we manufacture a search space using our Query Generator module, creating 19 possible Zoekt query candidates. The ultimate goal is trying to brute-force all the possible ways of constructing a Zoekt query, in order to maximize the chance of finding relevant code context. The key-word terms used to compose these queries are collected from the diff string associated with its completion point. This diff string can easily be constructed by comparing the concatenated string of *prefix* and *suffix* with the original code file before being modified by the revision. An example about the diff string could be found in Table V of the Appendix[4]. The diff string acts as the source of reference, so that we can use different **symbols gathering** techniques, utltilizing Tree-sitter for extracting meaningful identifiers from the located completion point. We start from function and class names, expanding it to navigation expression, and finally using all identifiers existing inside the diff string. The Tree-sitter syntax used for this process can be verified in Table VII of the Appendix.

To **construct Zoekt queries** from the gathered identifiers,

[4]Link to online Appendix https://doi.org/10.5281/zenodo.17045695

we used multiple types of combinations based on the official Zoekt query language guideline[5]. We start with naive exact matching, including all found identifiers in the queries, and gradually reduce the difficulty level of the query by reducing the number of terms inside the query, using regex for fuzzy search, and using OR logic to increase the search boundary. We ranked identifiers by their occurrence in the diff string and by their close proximity to the completion points, which is useful when filtering top-k identifiers. Table II shows the name and number of each query variation constructed in the public Kotlin and Python datasets. Examples of each query variation can be found in Table VI in the Appendix.

However, during online phase, not all query variations are sent to the Zoekt web server for searching. The Query Generator iterates on each variation, sending the generated query to the Zoekt web server, until it finds a non-empty search result. This ensures that we do not need to exhaustively search all query variations while the first few and most difficult variations have already successfully returned the results. We apply fallback, time-out and retry mechanisms to ensure that no requests are dropped when the server is overloaded. Another thing worth mentioning is that Zoekt allows Cross-shard searching, which means that queries can span multiple code revisions in the same repository. This can be done by omitting the specific revision ID in the query (Figure 3). This can be another way to increase the search coverage, similar to how developers searched through branches and commits. We experimented with both Single-shard and Cross-shard settings to see which settings have the highest successful search rate ( or hit rate). Results are reported in Table III.

Results returned from Zoekt contain metadata of the searched code snippets, including start, end lines and the containing file paths. The Post-processor module then uses this information to fetch the actual code snippets from the indexed shards.

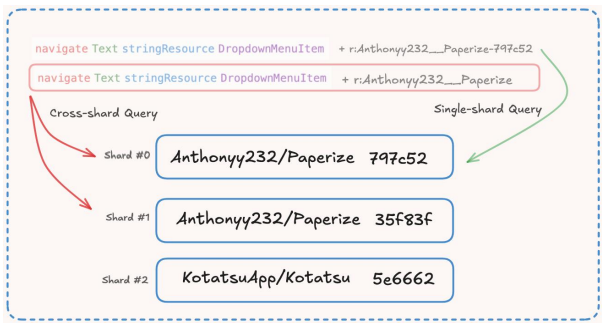

Fig. 3: Example of a Cross-shard Zoekt query spanning multiple code revisions in a repository.

*2) Post-processing of search results:* The Zoekt search API can return numerous results, to the extent of thousands of code snippets, especially in the case of Cross-shard querying. We first ranked them using Relevance Scores returned from Zoekt, and then find the containing file using the returned filepath.

[5]https://github.com/sourcegraph/zoekt/blob/main/doc/query_syntax.md

We use HuggingFace tokenizer to determine if the containing file is too large to be included in the final context window. If it does, we only return the code snippets or their union versions if they overlap with each other. We use Jetbrains Mellum's tokenizer [11] for both Python and Kotlin tokenization. At the same time, we implement a dynamic way to adjust the context window based on the size of the returned snippets. Specifically, we dynamically calculate the token constraint $T$ of each completion point by subtracting the maximum number of Mellum tokens from the total number of tokens existing in the diff string's prefix and suffix. A small token buffer is reserved to ensure there is enough token space for the CLMs to generate during the evaluation. Then, the post-processor selects up to top-k ranked files/snippets, ensuring each fits within a per-file token budget $R$ and the overall token constraint $T$, and concatenates their contexts for output (Equation 1 in the Appendix).

## III. FINDINGS

### A. Hit Rate between Single-shard and Cross-shard Setting

We define a "hit" as a response from Zoekt server which contains at least one successful search result. From Table III, the hit rate for Cross-shard setting was found to be significantly higher than that of Single-shard setting. This suggests that there could be more contexts located in other revisions of the same repository, which could be useful for code completion. This is analogous to how developers often search through branches and commits to find previous examples of implementations. We also investigated special cases where a query failed in the Single-shard Setting but succeeded in the Cross-shard Setting, and found that there was a high percentage of local contexts (found in the same containing file as the completion point) existing in other revisions. (Figure 4 in the Appendix)

|  | Single-shard Query | Cross-shard Query |
| --- | --- | --- |
| Python Hit | 213 | 239 |
| Python Miss | 34 | 8 |
| Kotlin Hit | 344 | 390 |
| Kotlin Miss | 56 | 10 |

TABLE III: Hit and Miss counts for Single-shard and Cross-shard settings in Python and Kotlin.

### B. Results in Public and Private set using Cross-shard setting

Since cross-shard queries achieve a much higher hit rate, we used submitted contexts found by this setting to the organizer's evaluation server for both the Public and Private phases. The results are shown in Table IV. Our solution ranked first in the Public Phase for Kotlin, reaching a chRF score of 0.7125. The Python public submissions also earned 0.6152, which qualified for second position. These optimistic results can be attributed to the high hit rate in Cross-shard queries (97.5% compared to 96.7% for Python), which leads to more relevant contexts being retrieved for code completion. For the Private Phase, we cannot confirm this assumption, as we do not have access to the evaluation data or hit rate information for that phase.

However, the results in the Private Phase showed a similar trend, with our submission outperforming every other in the Kotlin track. At the same time, the Python private submission ranked second, only slightly behind the first solution that uses semantic search.

It is also worth noting that searching through all 400 Kotlin public revisions took 18 minutes, while processing all 247 Python revisions required 12 minutes. Each retrieval request used a default timeout of 0.2 seconds when catching overloaded server exceptions. All of our results are collected on a Macbook M3 Air, using only the default container configuration of 1 CPU and 8GB RAM. This result demonstrates that our solution is lightweight but powerful in terms of performance, making it suitable for integration into existing in-IDE CLM-based code completion systems.

| Language | Public Phase | Private Phase |
|---|---|---|
| Kotlin | 0.7125 | 0.748 |
| Python | 0.6152 | 0.725 |

TABLE IV: chRF scores for Kotlin and Python in Public and Private phases.

## IV. DISCUSSION

### A. Threats to Validity

The effect of contexts found by Single-shard towards the quality of code completed is not investigated, leading to selection bias. There was no proven correlation between the hit percentage with the quality of code completion in a same programming language setting. Also, in the online retrieval phase, only the contexts are collected. No information about the effect of which specific symbols gathering techniques or Zoekt query construction variations on the search results is provided, which could limit the understanding of their contributions to the overall performance. Cross-shard setting can be considered as a threat to validity, as no information about the revision order is taken into account. In other words, a code snippet from a later revision might be retrieved to complete a code snippet from an earlier revision, which could be considered data leakage. Further data preparation steps are needed to mitigate this issue, and to prevent SpareCodeSearch from looking for contexts in future repository snapshots.

### B. Future Work

Apart from addressing the aforementioned threats to validity, our work opens new avenues for research. Trying to explore this solution in the context of secure code generation is a promising avenue. A prior work has used regex as Context Retrieval module [12] to localize Bug context used for automated patch generation with CLMs. Second, while our current implementation uses a fixed set of 19 query candidates for tractability, the space of all possible query formulations is significantly larger and potentially indefinite when considering different identifier combinations, regex patterns, and logical operators. Automated query optimization using search-based software engineering techniques [13] could explore this broader search space, discovering novel query strategies beyond our current hand-crafted variations.

IDEs and plugins developers could adapt SpareCodeSearch into their existing automated code completion features at ease. The Query Generator could make use of currently available code and AST parsing tools. Zoekt can be hosted entirely on a consumer laptop, and cross-shard setting could also benefit from temporal contexts extracted from version control tools. We publish SpareCodeSearch on Github[6], and welcome every contributions from the community to improve, extend it or to replicate our solution on other datasets and competitions.

---

[6]https://github.com/SPARE-UCD/spare-code-search